\documentclass[12pt]{article}
\usepackage[margin=1in]{geometry}
\usepackage{amsmath,amssymb}
\usepackage[round]{natbib}
\usepackage{hyperref}
\usepackage{setspace}
\usepackage{graphicx}
\usepackage{booktabs}
\hypersetup{
    colorlinks=true,
    linkcolor=blue,
    citecolor=blue,
    urlcolor=blue
}
\title{Rethinking the Scientific Method: \\An Introduction to Bayesian Epistemology}
\author{%
\parbox{\textwidth}{\centering
Jordan Hart\textsuperscript{1},
and Daniel W Franks\textsuperscript{1,2*}.
\\[0.5em]
\small
\textbf{1} Causa Ltd, York Hub, Popeshead Court Offices, York, UK, YO1 8SU \\
\textbf{2} Department of Biology, University of York, York, UK, YO10 5DD \\[0.5em]
\footnotesize
jordan@causa.tech (JH), dan@causa.tech, daniel.franks@york.ac.uk (DWF) \\
* Corresponding author
}}
\date{\footnotesize 10th July 2026}
\begin{document}

\maketitle

\begin{abstract}
Scientists routinely disagree not about data but about how to interpret evidence, because they implicitly operate from different epistemological frameworks without recognising it. The two dominant traditions, confirmationism and falsificationism, each capture genuine insights about scientific reasoning but face well-documented limitations. Confirmationism provides a natural account of how evidence supports hypotheses but cannot escape the problem of induction. Falsificationism provides logical rigour through deductive refutation but is undermined by the Duhem-Quine problem and offers no account of how scientists rationally accept theories and act on them. Here we argue that Bayesian epistemology provides a practical resolution to this impasse. By treating evidence probabilistically and operating over a finite, revisable set of hypotheses, the framework recovers the valid contributions of both traditions while addressing their core weaknesses. We show that confirmation and falsification emerge as special cases of Bayesian updating, that the subjectivity objection to priors is weaker than commonly supposed, and that the framework has direct practical consequences for study design, evidence synthesis, and publishing norms. Specifically, it replaces the falsifiability criterion with the more useful question of whether a hypothesis makes predictions that discriminate between competitors, and reframes the reproducibility crisis as an epistemological rather than a purely statistical problem. Adopting Bayesian epistemology, even informally as a mental model, can reduce friction between researchers, improve research efficiency, and help restore the cumulative character of scientific progress.
\end{abstract}

\section{Introduction}

Every time a scientist designs an experiment, interprets a result, or reviews a manuscript, they are implicitly invoking a framework for what counts as good evidence and sound reasoning. A reviewer who insists that a hypothesis must be falsifiable is channelling Popper \citep{popper1959}. A colleague who asks whether the data support the hypothesis is thinking like a confirmationist \citep{hempel1945}. A third who demands $p < 0.05$ is applying a convention whose philosophical underpinnings most practitioners have never properly examined, but that is different from falsification \citep{wasserstein2016}. These are not merely stylistic preferences; they reflect fundamentally different assumptions about how evidence relates to knowledge.

When collaborators and reviewers operate from different epistemological assumptions without recognising it, the result is not productive disagreement but unresolvable friction. Debates about whether observational studies ``count'' as evidence, whether null results are publishable, whether evidence supports one hypothesis over another, or whether a theory that cannot be easily falsified (such as certain formulations of string theory or evolutionary psychology) qualifies as science, are often epistemological disputes. Because the underlying philosophical disagreements are rarely identified, these debates consume time and resources without reaching a resolution.

The tools to resolve these problems already exist. Decades of work in philosophy of science and probability theory have produced a mature framework for reasoning about evidence \citep{carnap1950, lakatos1970, sprenger2019, jaynes2003}, but this work remains in specialist literature, often written for specialists. What has been missing is a clear account of how these ideas apply to the everyday decisions scientists face: which studies to run, how to weigh evidence, and how to choose between competing theories.

Bayesian epistemology offers a practical resolution to the problems outlined above. It is a single framework that accommodates the valid insights of both falsification and confirmation while providing clear guidance for working scientists. It does this through two key moves: first, treating evidence probabilistically via Bayes' theorem, so that each piece of evidence updates our degree of belief in competing hypotheses by a quantifiable amount; and second, requiring an explicit, finite set of plausible hypotheses with stated prior probabilities, while remaining open to expanding that set as science progresses. It recovers the valid contributions of falsificationism and confirmationism within a more general probabilistic framework while addressing their classical weaknesses. A shared practical epistemological framework would reduce friction between researchers, accelerate scientific progress, and open new avenues of discovery by making it clear which evidence to gather next. The implications are wider still: wherever evidence informs decisions, from policy and medicine to business and economics, the same reasoning applies.

This paper aims to raise awareness of the Bayesian epistemological framework among working scientists, both complementing and contrasting classical frameworks such as falsificationism, confirmationism, inductivism, and abductivism. The focus throughout is intented to be entirely practical: what mental models scientists should take away, what the framework means for study design and evidence evaluation, and where future work can operationalise these ideas. Section~\ref{sec:current} offers a characterisation of the dominant frameworks, confirmationism and falsificationism, and identifies their classical weaknesses. Section~\ref{sec:be} introduces Bayesian epistemology and section~\ref{sec:objections} addresses its principal objections. Section~\ref{sec:subsumes} shows that both traditions emerge as special cases of Bayesian Epsitemology. Section~\ref{sec:implications} draws out implications for study design, model selection, and publishing norms. Section~\ref{sec:howto} provides actionable guidance for adopting the framework, both as a mental model and as a formal method. Section~\ref{sec:conclusion} concludes with future directions.

\section{Current models of scientific reasoning}
\label{sec:current}

The diverse positions on scientific reasoning can be usefully organised into two broad traditions: \textit{confirmationism} and \textit{falsificationism}. This characterisation simplifies a complex and active academic field, but the dichotomy captures the essential tension that most practicing scientists navigate implicitly.


\subsection{Confirmationism}
\label{sec:confirmationism}

The first broad tradition holds that the goal of science is to find evidence that supports or confirms hypotheses. This is how many scientists implicitly work: they observe regularities, formulate hypotheses, and gather further evidence.  Also, what if you cannot falsify at all, yet you can collect evidence that is more consistent with one hypothesis than another? Can we not learn something still?
The tradition encompasses classical inductivism (reasoning from particular observations to general laws), inference to the best explanation (selecting the hypothesis that best accounts for the phenomena) \citep{lipton2004}, and Hempel's confirmation theory (seeking formal conditions under which evidence confirms a hypothesis) \citep{hempel1945}. \cite{carnap1950} attempted to place these intuitions on rigorous probabilistic foundations. What unites them is the direction of inference: from evidence to hypothesis acceptance.

Darwin's theory of natural selection is accepted because it best explains the diversity and adaptation of organisms. Some might say it is because it has survived severe attempts at falsification. But in reality it has (rightly) evolved as a theory as gaps between the theory and evidence have been found, and new ideas synthesized into the theory (e.g. genetics). When scientists review the literature, they routinely describe previous findings as `evidence for' or `support for' a hypothesis, reflecting a fundamentally confirmationist orientation.

Despite its intuitive appeal, confirmationism rests on a logical foundation that has been known to be unstable since Hume: the problem of induction \citep{hume1874philosophical}. No number of confirming observations logically guarantees a universal generalisation. The inferential step from ``$H$ predicts $E$'' and ``$E$ is observed'' to ``$H$ is supported'' is logically invalid: the evidence is consistent with $H$, but it may be equally consistent with other hypotheses. Put formally, $P(H \mid E)$ is not the same as $P(E \mid H)$. Goodman's ``grue'' paradox \citep{goodman1955} demonstrates that any finite set of observations is compatible with indefinitely many hypotheses that the data cannot distinguish between.

More sophisticated versions of confirmationism have been proposed, from Carnap's degree-of-confirmation functions to modern accounts of inference to the best explanation, but the core vulnerability persists. Confirmationism lacks a principled mechanism for weighting hypotheses against each other and cannot formally specify how much evidence is ``enough'' to accept a hypothesis. It leaves scientists with intuition where they need rigour.

\subsection{Falsificationism}
\label{sec:falsificationism}

The second broad tradition inverts the logic: rather than seeking evidence that confirms a hypothesis, it demands that scientists attempt to refute their hypotheses through severe falsification tests. Originating with \cite{popper1959}, falsificationism holds that a hypothesis is scientific if and only if it is falsifiable, meaning that there exist potential observations that would refute it. Importantly, falsification is distinct from null hypothesis significance testing (NHST) although it is often incorrectly conflated with it [box 1]. In its purest form, falsification proceeds by deductive logic: if hypothesis $H$ predicts observation $E$, and $\neg E$ is observed, then $H$ must be false. A failed prediction logically refutes its premise. This becomes difficult in practise when E is not binary like this. Hypotheses that survive rigorous attempts at falsification are ``corroborated'' but never confirmed or rendered probable. \cite{lakatos1970} refined this into a theory of research programmes, recognising that scientists do not abandon theories at the first anomaly but instead modify auxiliary hypotheses within a ``protective belt'' around the theory's hard core.

\textbackslash{}begin\{tcolorbox\}[title=NHST and p values]

Publication bias toward ``significant'' results distorts the scientific record \citep{ioannidis2005}, and reproducibility crisis has eroded public trust in science. A landmark replication effort found that only 36\% of psychology replications achieved statistical significance, compared to 97\% of the originals \citep{osc2015}. Over-reliance on null hypothesis significance testing (NHST) creates conditions conducive to well-documented pathologies including $p$-hacking \citep{simmons2011false}, HARKing (Hypothesising After Results are Known) \citep{kerr1998harking}, and the suppression of informative null findings \citep{wasserstein2016}. These are not merely statistical problems; they are epistemological ones. The way we think about evidence shapes what evidence gets produced and preserved. NHST, built around p-values, is explicitly asymmetric. You can only reject the null hypothesis or fail to reject it. ``Failing to reject'' is not the same as confirming evidence of the null, and it certainly does not provide evidence for your alternative hypothesis. For example, a small p-value says nothing directly about the probability that your theory is true (or false).

Most importantly, in practice, scientists recognise that multiple competing hypotheses can coexist, and that testing them in isolation is a limitation rather than a virtue. Consistency between observed data and the predictions of a hypothesis -- for instance, through a model -- constitutes evidence in favour of that hypothesis. This is not a methodological compromise; it has philosophical backing in likelihood reasoning and Bayesian inference, both of which treat evidence as a matter of degree rather than a binary verdict. The question is not whether we can confirm, but how honestly we acknowledge that we are doing so.

\textbackslash{}end\{tcolorbox\}

Falsificationism's appeal lies in its apparent logical rigour and its clean demarcation criterion: science is what can be falsified. Scientists frequently invoke this standard, asking ``but is it falsifiable?'' as the first test of whether a claim deserves to be taken seriously. The logical asymmetry between confirmation and refutation gives falsificationism its force: while no number of white swans can prove that all swans are white, a single black swan suffices to refute the claim. Of course, this common example is contrived and science rarely works with the logical convenience of  a single observation able to falsify a theory.

The most serious challenge to falsificationism is the Duhem--Quine problem \citep{duhem1954, quine1951}. Any empirical test involves not just the target hypothesis but a web of auxiliary assumptions about instruments, initial conditions, and background theories. When a prediction fails, the failure could lie in any component of this web. The refutation does not apply cleanly: we cannot tell whether the hypothesis $H$ or an auxiliary assumption $A$ is at fault; logically, we can only conclude that either $H$ \textit{or} $A$ is wrong. Scientists routinely ``save'' hypotheses by revising auxiliary assumptions, and sometimes they are right to do so. For example Le Verrier postulated Neptune to save Newtonian mechanics from the anomalous orbit of Uranus \citep{lakatos1978methodology}.

An equally important limitation is that falsificationism provides no formal mechanism for accepting hypotheses. Yet accepting and acting on hypotheses is precisely what practicing scientists do. Popper's ``corroboration'' is explicitly not a probability and is not meant to guide rational belief or decision-making. In practice, however, when a hypothesis has survived severe testing, scientists treat it as practically true: they build theories on it, design treatments around it, and make policy recommendations based on it. Yet falsificationism does not logically support these actions, leaving a consequential gap between the philosophy and the practice. 



\subsection{The impasse}
\label{sec:impasse}

Confirmationism and falsificationism arrive at a symmetrical impasse. Confirmationism provides a natural account of how evidence supports hypotheses but lacks logical rigour: it cannot avoid the problem of induction and offers no principled way to discriminate between hypotheses that equally accommodate the evidence. Falsificationism provides logical rigour through deductive refutation but is undermined by the Duhem--Quine problem and offers no account of how scientists rationally accept theories and act on them. Each captures something genuine about scientific reasoning, but neither captures the whole.

Without a shared framework, scientists risk wasted effort on studies that do not discriminate between hypotheses, publish results that cannot be meaningfully compared, and talk past each other in peer review. What is needed is not a third competing framework but a practical one that preserves the confirmationist insight that evidence genuinely supports hypotheses while respecting the falsificationist insistence on rigorous testing and the provisional status of all knowledge. This requires two things: a principled, formal way to interpret evidence as bearing on hypotheses, both for and against; and a willingness to operate with a finite set of plausible hypotheses at any given time while remaining open to expanding that set as science progresses. Bayesian epistemology achieves exactly this.

\section{Bayesian epistemology}
\label{sec:be}

Bayesian epistemology represents a unified approach to scientific inference. Instead of dealing with absolutes (hypotheses are true or false), it deals with probabilities. Hypotheses are assigned probabilities that are updated over time as new evidence becomes available. This means that all available information can be brought together to improve the state of scientific knowledge at a point in time, and can be built upon as new information becomes available. This also means that disparate information of different types and sources can be pieced together to form a richer understanding. This approach captures the desirable properties of both confirmation and falsification, while avoiding each of their major pitfalls.

The subsections below outline the mathematical foundations of Bayesian epistemology, describe how they form a framework for scientific inference, and give a real world example.

\subsection{Probability as degree of belief}

At the foundation of Bayesian epistemology lies a particular interpretation of probability: probability represents a rational agent's degree of belief in a proposition, calibrated and updated in light of evidence \citep{earman1992, howson2006}. This contrasts with the frequentist interpretation of probability, in which probability refers specifically to the long-run relative frequency of an outcome after many repeated trials. A key difference is that Bayesian probability gives the probability of a hypothesis given evidence, whereas frequentist probability does the opposite. Under the Bayesian interpretation, it is meaningful to assign a probability to a singular hypothesis, such as ``dark matter is composed of weakly interacting massive particles,'' even though there is no repeatable experiment to which a frequency interpretation could usefully apply \citep{jaynes2003}.

\subsection{Updating beliefs from evidence}

By assigning probabilities to hypotheses, Bayes' theorem can be used to update beliefs as new evidence arrives. For a hypothesis $H$ and evidence $E$, the probability of the hypothesis after seeing the evidence, the \textit{posterior probability}, is given by:
\begin{equation}
\label{eq:bayes}
P(H \mid E) = \frac{P(E \mid H)\, P(H)}{P(E)}
\end{equation}
where $P(H)$ is the \textit{prior probability}, our degree of belief in $H$ before observing $E$; $P(E \mid H)$ is the \textit{likelihood}: how probable the evidence is if $H$ is true. $P(E)$ is the \textit{marginal likelihood}, the total probability of the evidence under all hypotheses. 

The learning rule is simple: upon finding evidence $E$, the probability of the hypothesis $H$ can be updated to $P(H \mid E)$ \citep{earman1992}. However, in isolation, $P(E)$ is not known and therefore the posterior probability cannot be calculated. Fortunately this can be overcome by making a minimal but key assumption.

\subsection{Probabilities over multiple hypotheses}

If there are only a finite number of hypotheses being considered, $P(E)$ no longer needs to be calculated directly, due to the law of total probability. To show this, consider a finite set of hypotheses $H^* = \{H_1, H_2, \ldots, H_n\}$, alongside prior probabilities $P(H_1), P(H_2), \ldots P(H_n)$, then Bayes' theorem provides a method to calculate posterior probabilities for each of the hypotheses:

\begin{align}
p(H_i \mid E) &= \frac{p(E \mid H_i)\, p(H_i)}{p(E)} \nonumber \\[6pt]
              &= \frac{p(E \mid H_i)\, p(H_i)}{\sum_j^N p(E \mid H_j)\, p(H_j)}
\label{eq:multiple-hypotheses}
\end{align}

This makes it possible to hold a set of initial beliefs over a set of hypotheses, observe new evidence, and update beliefs over the entire set of hypotheses at once. Each piece of evidence shifts beliefs in proportion to how well it discriminates between hypotheses. Evidence that is equally expected under all hypotheses is uninformative and does not change the posterior. Evidence that is much more probable under one hypothesis than another shifts beliefs strongly. 

This is the kind of reasoning many scientists generally already do informally or implicitly, but Bayesian epistemology makes it more explicit, formal, precise, and actionable \citep{sprenger2019}.

\subsection{Scientific inference with Bayesian epistemology}

Equation \ref{eq:multiple-hypotheses} gives a formal mathematical mechanism for updating beliefs based on evidence, but Bayesian epistemology is a broader framework that describes how to use this equation for scientific inference. The framework follows these steps:
\begin{enumerate}
    \item Specify a set of plausible hypotheses with prior probabilities reflecting current background knowledge.
    \item Design studies that best discriminate between viable hypotheses.
    \item Run the studies and collect evidence.
    \item Update posterior probabilities using equation \ref{eq:multiple-hypotheses}. These become the priors in the next iteration.
    \item If new hypotheses emerge, add them to the set.
    \item Iterate (go to 2).
\end{enumerate}

This is not a one-shot procedure, but an ongoing and self-correcting cycle in which each round of evidence refines our understanding and informs the design of subsequent investigations. The cycle is efficient by design: at each step, resources are directed toward the evidence that will be most informative, rather than toward studies whose outcome would leave beliefs unchanged.


\subsection{Example: Extinction of the dinosaurs}

To illustrate how the Bayesian epistemology framework can work in practice, we consider a much simplified example concerning the causes of the Cretaceous–Paleogene extinction event that resulted in the extinction of all non-avian dinosaurs. As recently as the 1970s, several serious hypotheses to explain this had been proposed and were competing for scientific support. The history of research in this area is fascinating and complex in its own right. But our example is not a serious study of the historical timeline of research activity in this area, simply an illustrative example of how Bayesian epistemology can be used in science.

For simplicity we will assume that prior to 1980, the key hypotheses were: $H_A$: climate cooling; $H_B$: volcanic activity; $H_C$: sea-level regression; $H_D$: impact event; and $H_E$: other hypotheses.

Prior to the discovery of high levels of iridium (a consequence of an ``extra-terrestrial'' impact event) in the Cretaceous–Paleogene boundary by Alvarez et al. in 1980, the impact event hypothesis was a relatively fringe theory. The predominant theories at the time were related to climate change, volcanic activity, and sea-level regression, all of which were supported to varying degrees by evidence. Following the discovery of the high iridium boundary, it became clear that an impact event occurred around the same time as mass extinction. The state of knowledge at this point can be translated into the Bayesian epistemology framework through priors that were held across the hypotheses, with the concentration of prior probability around hypotheses $H_A$, $H_B$, and $H_C$.

Upon discovery of evidence of impact $E$, we only need to assign a probability of seeing such evidence given each hypothesis to calculate the posterior probability. For example, for hypothesis $H_A$, the probability of the new impact evidence if climate cooling was the true cause of extinctions is equal to the background probability of an unrelated, independent impact event around the same time as the extinction. In this example we will assume this probability is 1\%, though it is likely far lower. This gives our likelihood for all hypotheses except for the impact event $H_D$, where the evidence is expected exactly under the hypothesis, so it receives a likelihood of $1.0$. It is important to note that this is not yet the posterior probability of the hypothesis, but the likelihood of the evidence given the hypothesis is true. Using equation \ref{eq:multiple-hypotheses}, we can calculate the posterior probabilities shown in Table \ref{tab:dinosaur_example}.

The updated probabilities in light of the impact evidence show that the impact event hypothesis $H_D$ now has a much higher credence than before at around 84\%, and the former leading hypotheses are downgraded to around 6.7\% (uncertainty is also meaningfully quantified). This process can be repeated as more evidence becomes available to further refine the posterior probability estimates. This process matches closely with what actually happened: after the initial discovery of impact evidence, scientific belief in the theory quickly increased, but it took several years of further evidence before the hypothesis was fully accepted by scientific consensus.
\begin{table}[h]
\centering
\begin{tabular}{lcccc}
\toprule
\textbf{Hypothesis $(H_k)$} & \textbf{Prior} & \textbf{Likelihood $P(E \mid H_k)$} & \textbf{Posterior} \\
\midrule
Climate Cooling ($H_A$) & 0.40 & 0.01 & 0.067 \\
Volcanic Activity ($H_B$) & 0.10 & 0.01 &  0.017 \\
Sea-level Regression ($H_C$) & 0.40 & 0.01 &  0.067 \\
Impact Event ($H_D$) & 0.05 & 1.0 &  0.840 \\
Other ($H_E$) & 0.05 & 0.05 & 0.009 \\
\bottomrule
\end{tabular}
\caption{Bayesian updating with five competing hypotheses about the cause of the extinction of the dinoaurs. The likelihood is assigned based on the probability of seeing an impact event around the same time as the dinosaurs went extinct, either by coincidence or in a causative sense, depending on the hypothesis. Despite the impact event hypothesis $H_D$ having low prior probability, its predictions match so well with the evidence while the other hypotheses do not, that it gains a large amount of probability weighting in the posterior.}
\label{tab:dinosaur_example}
\end{table}

It is also worth noting that evidence against competing hypotheses would have served a similar role as falsification. For example, if concrete evidence was found that the climate actually warmed over this period, the likelihood of such evidence given the climate cooling hypothesis $H_A$ would have been near zero, which would relegate hypothesis $H_A$ to a low posterior probability.

For many practicing scientists, this framework will feel familiar to how they practically do things; and that is the point. Scientists already weigh evidence, update their views, and consider multiple hypotheses, though they often do so informally opaquely \citep{kahneman2011}. Bayesian epistemology captures the logic of what good scientific reasoning already does. But formalising what scientists already do informally brings clarity, reduces inconsistency, and addresses longstanding problems \citep{jaynes2003, sprenger2019}.


\section{Addressing key objections}
\label{sec:objections}

\subsection{Subjectivity of priors}

The most frequently raised objection to Bayesian epistemology and Bayesian thinking more generally is that prior probabilities are subjective: if different scientists start with different priors, will they not reach different conclusions? \citep{fisher1925statistical,goldstein2006subjective}


The answer is yes, and that is a desirable property. Priors are not arbitrary, and refusing to state them is not the same as not having them. A framework that declines to assign prior probabilities is implicitly treating all hypotheses as equally plausible before the data arrive, an extraordinarily strong assumption that no practicing scientist usually holds. No serious researcher considers a hypothesis with no theoretical motivation, no mechanistic basis, and no prior empirical support to be as plausible as one grounded in decades of converging support (even anecdotal or theoretical). The question is not whether scientists have priors but whether they make them explicit. When priors are explicit, a scientist who assigns implausible priors is making a substantive claim that can be scrutinised and corrected. A framework that pretends to have no priors at all is making an even stronger claim while evading scrutiny entirely. In the rare situation where multiple hypotheses are considered as likely, equal weight can be assigned in priors anyway, if that is desirable. But again, that choice is then explicit and can be scrutinised. Hiding assumptions is not the same as not having them.

It is worth noting that the likelihood function $P(E \mid H)$, which frequentist and Bayesian methods alike use, is itself a modelling choice. Just like priors are. The likelihood function depends on distributional assumptions, on what counts as signal versus noise, and on how covariates are parameterised. These choices are every bit as contestable as the choice of prior, yet critiques that likelihoods are ``subjective'' are extremely rare. The difference is psychological: the likelihood feels objective because it concerns the probability \textit{of the data} rather than the probability \textit{of the hypothesis}. A critic who objects to priors on grounds of subjectivity but accepts a likelihood function without scrutiny is applying an inconsistent standard.

\subsection{Science requires falsifiability}

A frequent objection to certain kinds of study designs is that ``science requires falsifiability'' and that non-falsifiable designs are not scientific and should be treated with suspicion at best, and derision at worst. This position has its roots in Popper's answer to the \textit{demarcation problem}, which seeks to formally distinguish between science and non-science \citep{popper1959}. Yet in many important cases, falsifiability is neither possible nor practical. Yet understanding can still progress. Returning to our extiction example, it is impossible to truly falsify the impact event hypothesis. Yet evidence in favour has swung scientific consensus.


Bayesian epistemology overcomes this logical loophole by simply assigning extremely low probabilities in places where hypotheses would otherwise be said to be ``falsified''. This retains the desirable properties of falsification while providing a pragmatic way forward. It also means that non-falsifiable study designs should be treated rigorously, and where their non-falsifiability restricts information gain, their influence will be down-rated. Where they provide a lot of new information, their influence can be significant.

\subsection{Infinite hypotheses}

The Bayesian epistemology framework requires that the set of plausible hypothesis considered at any one time be restricted to a finite set. A critique of this requirement is that the world of hypothesis could be infinite. Through induction it is always possible to modify a hypothesis slightly and generate a new hypothesis. This process could be continued to provide an infinite world of competing hypotheses \citep{hume1874philosophical}.

There are two elements to the framework that prevent this from occurring: 1) a hypothesis in the set must be \textit{plausible}, unrealistic or tortured hypotheses should not be included in the set (their prior probability rules them out); and 2) the finite set can change and be added to over time. The definition of plausibility may be debated between scientific peers. This is not a disadvantage, but a strength of the Bayesian epistemology framework. By agreeing on shared parameters and particular points of disagreements, it enables such debates to be fruitful. 

It is also natural that new ideas and new discoveries should generate previously unconsidered hypotheses, and if deemed plausible these should be added to the set too. This shares themes with Kuhn's idea of ``scientific revolutions'' in which new ideas can completely overturn existing understanding \citep{kuhn1962}. In Bayesian epistemology, if a new hypothesis is introduced with far more evidence, all other hypotheses will be heavily down-graded and the new hypothesis will become favoured. The finite nature of the plausible set simply represents the fact that only a finite number of hypotheses are being considered at any single point in time, it does not prevent new ones from being added.

\section{Bayesian epistemology recovers other frameworks}
\label{sec:subsumes}

By ``recovers'' we mean that the earlier insight can be expressed within the Bayesian framework.

\subsection{Recovering confirmationism}

The central claim of confirmationism is that repeated observations supporting the predictions of a hpothesis can rationally increase our confidence in a hypothesis. This core insight emerges as a natural consequence of the Bayesian epistemology framework, while resolving its core logical vulnerability. By replacing deductive confirmation with probabilistic updating, it is possible to increase credence in a hypothesis after repeatedly observing evidence congruent with it. Observing evidence $E$ predicted by hypothesis $H$ does not ``confirm'' $H$ in any deductive sense. Instead, $E$ raises $P(H)$ by exactly the amount warranted by the likelihood ratio $P(E \mid H) / P(E \mid \neg H)$. If another hypothesis $H'$ predicts $E$ equally well, $E$ does not discriminate between them and the posteriors are unchanged relative to each other. This is the formally correct version of the intuition that confirmationism tries to capture, freed of its logical fallacy. The 'ravens paradox' is often used as a criticism of confirmationism. But the criticism does not stand here when it receives a natural Bayesian treatment: observing a white shoe provides at most negligible evidence that all ravens are black, because the observation is approximately equally expected whether or not all ravens are black \citep{howson2006, earman1992}. A black raven is far more informative because ravens are rare compared to non-black things.

\subsection{Recovering falsificationism}

Falsification, rather than being an alternative to Bayesian epistemology, emerges as a special case: when evidence is highly improbable under a hypothesis, Bayesian updating drives that hypothesis's posterior probability toward zero. If $H$ predicts $E$ with near-certainty and $\neg E$ is observed, $P(H \mid \neg E)$ drops dramatically. This is the Bayesian analogue of deductive refutation, but probabilistic rather than the all-or-nothing falsification approach. The key advantage is that it accommodates degrees: a mildly surprising result reduces a hypothesis's posterior probability mildly, while a starkly contradictory result reduces it dramatically. This is more realistic than all-or-nothing falsification, which must either declare a hypothesis dead or (commonly) invoke auxiliary hypotheses to save it, with no formal guidance for choosing between these options \citep{howson2006, sprenger2019}.

The concept of severe testing, the idea that hypotheses are most strongly supported when they survive tests that had a high probability of revealing them to be false, has a direct Bayesian epistemology interpretation \citep{mayo1996, mayo2018}. A severe test is one where the evidence strongly discriminates between hypotheses: $P(E \mid H)$ is very different from $P(E \mid \neg H)$. In Bayesian terms, this corresponds to a high likelihood ratio. A hypothesis that survives a test where the likelihood ratio is 100 receives a much larger posterior boost than one where the ratio is only 2.

The Duhem--Quine problem is not eliminated by Bayesian epistemology, but instead is explicitly captured by it. When a prediction fails, posterior probability is redistributed across all hypotheses in proportion to their prior plausibility and their likelihood of generating the observed result. If the auxiliary assumptions are well-established, carrying high prior probability from extensive independent evidence, most of the ``blame'' falls on the target hypothesis. If the auxiliaries are themselves uncertain, the blame is distributed more evenly \citep{dorling1979, strevens2001}. It is even possible to update the probabilities over auxiliary assumptions if desired.



\subsection{Why choose Bayesian epistemology?}


Like any approach to science, Bayesian epistemology is not free of weaknesses. But its weaknesses are the transparent counterparts of weaknesses that other frameworks harbour invisibly and more strongly. It is not a compromise between confirmationism and falsificationism but a more expressive formal language that recovers the valid insights of each. It captures both the confirmationist insight that evidence supports hypotheses at the same time as it captures the falsificationist insight that severe tests are more informative. The provisional status of all knowledge is built into the framework: no hypothesis ever reaches posterior probability 1. What Bayesian epistemology adds is what neither classical tradition provides: a single rigorous framework for weighing evidence, whether it supports or undermines a hypothesis \citep{sprenger2019, howson2006}.

Crucially, this is not a call to replace one dogma with another. In principle, no hypothesis is ever beyond revision within the Bayesian framework, no evidence is dismissed on procedural grounds, and every assumption is stated openly enough to be challenged. In practice, very strong priors combined with moderate evidence can entrench beliefs. But the framework makes this entrenchment transparent: the priors are stated, the sensitivity to them can be checked, and anyone can see whether the posterior is driven by evidence or by prior conviction. This transparency is itself a safeguard against dogmatism that other frameworks lack. The same structure is what makes the framework efficient: because every study, every observation, and every piece of prior knowledge feeds into the same updating process, resources are naturally directed toward the evidence that will shift understanding the most. The framework simultaneously tells scientists what to investigate next, how to interpret what they find, and when to change their minds.

\section{Broader implications for scientific practice}
\label{sec:implications}

\subsection{Priors are important, and should be explicit}

The choice of which hypotheses to consider, and what prior probability to assign each, is not a preliminary step to be glossed over but a central scientific judgment. As with any approach, if the true hypothesis is excluded from the set, no amount of data can identify it. If it is assigned a negligibly low prior, it will require overwhelming evidence to overcome its handicap. If too many implausible hypotheses are included, posterior mass is spread thinly. These are not weaknesses of the Bayesian framework; they are the visible form of choices that every approach to science makes, whether it acknowledges them or not \citep{jeffreys1946invariant,lemoine2019moving}.

Priors do not need to be precise to be useful. A prior that rules out physically impossible values, constrains effect sizes to plausible ranges, or assigns higher probability to mechanistically grounded hypotheses is doing substantive scientific work, even if the exact numbers are debatable \citep{gelman1995bayesian}. The sensitivity of conclusions to reasonable variation in priors can and should be checked: where conclusions are robust, they can be held with greater confidence; where they are sensitive, this itself is informative, revealing that the evidence is insufficient to overcome prior uncertainty.

The well-known problem of multiple comparisons takes a different form in the Bayesian framework: including many hypotheses dilutes posterior probability across candidates, but restricting the set to genuinely plausible candidates restrains this dilution \citep{gelman2016statistical}.

\subsection{Running the right studies}


The best experiment is the one that most strongly discriminates between competing hypotheses. \cite{lindley1956} formalised this as \textit{expected information gain}: the experiment that maximises the expected shift from prior to posterior. Bayesian epistemology lays the groundwork to be able to ask ``what is the single most informative piece of evidence we should obtain next?''.  This makes science more efficient and focused.

Randomised controlled trials (RCTs) are often held up as the ``gold-standard'' of scientific evidence \citep{deaton2018understanding}. However under Bayesian epistemology, there is no categorical hierarchy of evidence that necessarily holds RCTs as higher in value than, say, observational studies (e.g. with causal inference) or theoretical work that are well analysed. In fact, RCTs can fail in ways that observational studies do not \citep{cartwright2007rcts}. Therefore all evidence must contribute to the posterior, weighted by its reliability (via auxiliary assumptions) and informativeness. Randomised controlled trials, observational studies, meta-analyses, case reports, theoretical arguments, and simulation results all produce likelihoods that update posteriors. The key is that each source of evidence must be incorporated correctly \citep{berry2006, trotta2008}.


The Bayesian framework replaces the falsifiability criterion with a more useful one: a hypothesis is scientifically informative insofar as it makes predictions that differ from those of its competitors. The question shifts from ``can this hypothesis be falsified?'' to ``does this hypothesis make differential predictions?'' This respects the Duhem--Quine insight that no individual hypothesis is falsifiable in isolation \citep{earman1992}.




\subsection{Implications for the reproducibility crisis and publishing norms}

Viewed through a Bayesian epistemology lens, the reproducibility crisis discussed in Section~\ref{sec:impasse} is not merely a statistical problem but an epistemological one. Publication bias toward $p < 0.05$ means that the published literature over-represents evidence for hypotheses and under-represents evidence against them. If scientists can only access a biased subset of the evidence, any posteriors will be systematically miscalibrated. The scale of this miscalibration is striking: Berger and Sellke \citep{berger1987} demonstrated that $p$-values of 0.05 frequently correspond to posterior probabilities of the null hypothesis of 0.30 or higher. 


Bayesian epistemology suggests that every study, whether positive, negative, or ambiguous, contributes to the posterior and deserves to be part of the scientific record. Null results are not failures; they are evidence that updates posteriors, sometimes strongly. The current incentive structure, which rewards novel significant findings and punishes null results, is antithetical to good scientific reasoning. 


The reproducibility crisis is partly driven by institutional incentives that no purely individualist epistemology can fully capture. But Bayesian epistemology provides a framework within which these distortions are defined: publication bias is the systematic suppression of likelihoods; $p$-hacking is the distortion of likelihood functions; HARKing is the post-hoc construction of hypotheses with favourable likelihoods \citep{sprenger2019}. Bayesian epistemology provides a common framework to understand these issues, and make the case for their resolution.

\subsection{Accommodating paradigm shifts}

Kuhn's \citep{kuhn1962} account of scientific revolutions maps naturally onto the Bayesian framework. Within a paradigm, scientists perform Bayesian updating over an established hypothesis set. Anomalies accumulate as observations that fit poorly under all current hypotheses, signalling that the hypothesis set may need expansion. A paradigm shift occurs when a new hypothesis is articulated that explains these anomalies, rapidly accumulating posterior probability as the updating machinery redistributes belief.

\section{Practical guidance for using Bayesian epistemology}
\label{sec:howto}


The most important recommendation of this paper is also the simplest: adopt Bayesian epistemology as a mental model. When planning a study, the overarching question is: given what we currently believe, what is the single most informative thing we could investigate next? This breaks down into concrete sub-questions: what are my competing hypotheses, and what evidence would best discriminate between them? When reviewing a paper, ask: how much does this evidence shift my beliefs, and does the study design maximise information gain? When encountering a surprising result, ask: is this better explained by updating my model or by questioning an auxiliary assumption? In each case, the payoff is the same: getting closer to understanding what actually causes the phenomenon you are studying. These questions do not require Bayesian computation. They require Bayesian thinking.

One concrete change: stop using ``but is it falsifiable?'' as the default criterion for evaluating scientific work. Replace it with ``does this hypothesis make predictions that differ from those of its competitors?'' This question is more useful because it focuses on evidential content rather than an abstract logical property, more permissive because it allows theoretical work, computational modelling, and historical sciences to be evaluated on their merits, and more honest because it acknowledges the Duhem--Quine insight that no hypothesis is testable in isolation \citep{duhem1954, quine1951}.

Equally, resist the confirmationist reflex of treating agreement between a hypothesis and the data as strong evidence for that hypothesis. Agreement is only informative insofar as it discriminates: if all plausible hypotheses predict the data equally well, the data are uninformative regardless of how impressive the agreement appears. Always ask: which hypotheses does this evidence favour over which others, and which does it disagree with? 

Stop requiring binary dichotomisations (significant versus not significant, confirmed versus refuted) and instead reason in terms of degrees of evidence. Where feasible, report posterior probabilities, likelihood ratios, or credible intervals alongside or instead of $p$-values \citep{wasserstein2016, wagenmakers2007}. Encourage the publication of all results, including null results, as contributions to the cumulative evidence base. When evaluating a body of evidence, synthesise it rather than counting studies that crossed an arbitrary threshold.

Maintain humility about the completeness of your hypothesis set. The history of science is a history of surprises, and the practical lesson is attitudinal: when anomalies accumulate and no current hypothesis can account for them, resist the temptation to explain them away with ever more elaborate auxiliary hypotheses. Consider whether a genuinely new hypothesis is needed. This is the Bayesian account of Kuhn's \citep{kuhn1962} paradigm shifts, and it guards against the dogmatism that both confirmationism and falsificationism can encourage in practice.

A common objection to hypothesis testing from practitioners in exploratory fields is that they do not have a well-defined hypothesis set to begin with. This reaction is forced by the requirement of falsification in publication norms. Bayesian epistemology does not require starting with the right hypothesis set, only with \textit{a} hypothesis set that is iteratively refined. Exploration, from pilot studies to qualitative research to exploratory data analysis, is the process of building a hypothesis set good enough to test. The framework provides a clear account of what ``good enough'' means: a hypothesis set is adequate when the current hypotheses collectively explain the observed data well, as assessed by posterior predictive checks \citep{gelman2013}. Far from rejecting exploratory science, Bayesian epistemology provides a formal account of what exploration is \textit{for} and when it has succeeded.


Future work should seek to formalise Bayesian epistemology quantitatively. Meta-analyses are a natural fit: prior distributions encode results from previous studies, and the posterior synthesises the full evidence base \citep{hoeting1999}. Clinical trials increasingly use Bayesian adaptive designs that update evidence in real time and can be both more efficient and more ethical than fixed-sample designs \citep{berry2006}. Astrophysics has adopted a conceptually similar approach as standard methodology for model selection and parameter estimation \citep{trotta2008}.



The barrier to adoption is low. As a mental model, Bayesian epistemology requires no software, no new statistical training, and no institutional change. It requires only a willingness to ask: what are my hypotheses, what evidence would discriminate between them, and how should that evidence change my mind? As a formal method, there is much scope to build on existing methodologies such as Bayesian model comparison and Bayesian networks to rigorously implement this framework.

\section{Conclusion}
\label{sec:conclusion}

Bayesian epistemology provides a practical framework for scientific reasoning that recovers the valid insights of both confirmationism and falsificationism while addressing their classical weaknesses. Confirmationism captures how evidence supports hypotheses but faces the problem of induction. Falsificationism captures the importance of severe testing but is undermined by the Duhem--Quine problem and lacks a mechanism for rational acceptance. Bayesian epistemology preserves both insights: evidence supports hypotheses through posterior probability increase, severe tests produce large shifts in posterior probability, auxiliary hypotheses are handled through probability redistribution, and all conclusions remain provisional.

This paper aims to make the Bayesian epistemology mindset accessible. The natural next step is to build the tools to formally apply the framework. In most fields, hypotheses relate directly to causal phenomena: ``why does y happen?'', ``does x cause y?'', ``what would have happened to y if x'', and similar. Answering causal questions requires reasoning over competing causal structures. The Bayesian epistemology frameworks maps directly onto this: assign prior probabilities to competing structures, use expected information gain \citep{lindley1956, chaloner1995} to determine which experiments to run and where to collect data, update posteriors as evidence accumulates, and iterate. This approach has the potential to turn Bayesian epistemology from a conceptual framework into a concrete tool for scientific discovery, one that tells researchers not just how to interpret evidence but what evidence to gather next. A companion paper will develop this application in full.

Bayesian epistemology has implications far beyond the philosophical. Scientists who adopt this framework will design studies that discriminate more sharply between hypotheses, integrate evidence from diverse sources rather than discarding inconvenient data, and report results that tell readers how much beliefs should shift rather than whether an arbitrary threshold was crossed. The implications extend beyond science: any domain where evidence informs decisions, from clinical medicine and public policy to business strategy and economics, faces the same fundamental questions about how to weigh competing hypotheses, integrate diverse evidence, and act under uncertainty. Bayesian epistemology provides a coherent answer.


\newpage
\bibliographystyle{plainnat}
\bibliography{references}

\end{document}